# Monte-Carlo-based 4D robust optimization using physical and temporal uncertainties for intensity-modulated proton therapy

Mark D. Pepin,[a] Erik Tryggestad, Hok Seum Wan Chan Tseung, Jiasen Ma, Abdou Abdel Rehim, Jedediah E. Johnson, Michael G. Herman, and Chris Beltran
*Department of Radiation Oncology, Mayo Clinic, 200 1st Street Southwest Rochester, MN, 55905, USA*

**Purpose:** Respiratory motion and the interplay effect cause the dose delivered to a patient with spot-scanned proton therapy to differ from the dose planned for during optimization. A new 4D robust optimization methodology was developed which incorporates patient breathing and the interplay effect, as well as their uncertainties, into the optimization process to generate plans which are more robust against motion.

**Methods:** The 4D robust optimizer used a 4DCT image set to obtain information regarding patient breathing and included a beam-delivery simulation to incorporate the interplay effect. The doses from potential spots were calculated on all breathing phases using Monte Carlo simulation, deformed to a reference phase using deformable image registration, and then added in a weighted sum. The weights for the different phases were based on the beam-delivery simulation. Robustness was added to the optimization by considering range and setup uncertainties as physical-uncertainty scenarios and uncertainties in the patient breathing and treatment delivery as temporal uncertainty scenarios. Phase-weighted doses were calculated for each scenario and an all-scenario robust optimization engine used to derive final plans. Most primary components were accelerated by using graphics processing units, including the Monte Carlo and beam-delivery simulations, the dose deformations, and the optimization engine. Plans were generated on three test patients using the 4D robust optimizer and a 3D robust optimizer and the resulting plans compared.

**Results:** The 4D robustly optimized plans exhibited improved target coverage with an average increase in the dose to 98% of the target volume of 4.9%, improved homogeneity with an average decrease in the homogeneity index of 6.0%, and improved robustness with decreased ranges in coverage and homogeneity amongst all uncertainty scenarios by averages of 35.0% and 52.5%, respectively. Small increases in organ at risk maximum doses were observed, but the increases were within the input constraints for the optimizations.

**Conclusions:** A 4D robust optimizer was developed which included the interplay effect and did not depend on the synchronization of breathing and delivery, on their repeatability with each fraction, or on fractionated averaging. Plans generated showed improved dose characteristics compared to 3D robust optimization.

## 1. INTRODUCTION

While the conformal nature of proton therapy increases organ at risk (OAR) sparing,[1–3] it also increases the treatment's sensitivity to uncertainties affecting the localization of the deposited dose.[4,5] Intensity-modulated proton therapy (IMPT) is particularly susceptible to uncertainties,[6,7] as individual IMPT fields are inhomogeneous with sharp dose gradients such that relative shifts between and within fields would be more prone to create regions of over and under dosage. One method developed to guard against this sensitivity is robust optimization;[8] the dosimetric impacts of potential uncertainties are given as inputs to the inverse-optimization algorithm to generate less sensitive (*i.e.*, more robust) treatment plans. Each uncertainty is manifest as a "robust scenario" for which dose is evaluated and the optimization algorithm combines information from all such scenarios to determine the optimal treatment plan. Most robust-optimization algorithms in the literature focus on potential physical changes due to patient-setup and proton-range uncertainties.[7,9–13] However, for any treatment target near the diaphragm, target motion due to respiration introduces additional uncertainty which affects the resulting dose.[14,15]

Motion from patient breathing can displace tumors and organs in proximity to the lungs by up to 20 mm[16] and can cause loss of coverage and homogeneity in the delivered dose distribution. For IMPT delivered by spot-scanning, the so-called "interplay effect" creates an additional avenue for dose degradation.[17–20] The interplay effect arises from a lack of synchronization between patient breathing and dynamic spot delivery, resulting in dose degradation (non-uniformity), *i.e.*, sub-volumes or regions in the anatomy of constructive and destructive spot interference. There are several approaches which have been clinically used to reduce or mitigate the interplay effect[21]: deep-inspiration breath hold,[22] gated delivery,[23] repainted delivery,[24] and target tracking.[25] However, these are non-ideal solutions as the first three strategies can greatly increase treatment time and tumor tracking is technically challenging.

Given the development of robust optimization for standard (*i.e.*, non-motion) three-dimensional (3D) cases to reduce sensitivity to uncertainties, four-dimensional (4D) robust optimization is a logical mechanism to reduce the interplay effect without greatly increasing treatment time or requiring new equipment. The defining characteristic of 4D optimization is the use of multiple CT image sets in the optimization process; however, optimization approaches vary throughout the literature and are non-ideal for various reasons. The methods by Eley *et al.*[26] and Graeff *et al.*[27] utilized 4D optimization on top of target tracking, and are thus non-viable solutions for clinics without tracking capability. Some approaches considered optimizing on spot fluences which are breathing-phase or target-location specific.[27–29] This may result in ideal optimization solutions, but requires synchronization between breathing-tracking and beam-delivery control systems to realize this optimal solution during treatment and, therefore, would require increasing the complexity of the delivery-control system notwithstanding the concern over how to accurately monitor internal target motion. Liu *et al.*[30] took the step of considering robust 4D optimization, using the typical setup and range uncertainties, assuming that the dose from each spot was spread uniformly throughout the breathing cycle, *i.e.*, performing multi-image optimization (MIO) with the 4DCT image sets. This assumption is only valid in passive-scattered proton therapy or in IMPT with a large number of fractions such that the interplay effect is averaged out.[31] However, stereotactic body proton therapy (SBPT) is now in clinical use for lung,[32–34] liver,[35] and other abdominal tumors[36] and the very few fractions (approximately 3–10) used in such a treatment would break the assumption of uniform dose throughout the breathing cycle. Additionally, although the physical doses may average out over many fractions, there is a question of whether the inhomogeneous biological effects caused by each fraction would also average to a homogenous biological effect.[37]

To correctly consider optimization of IMPT treatments without synchronization or fraction averaging, information about the beam delivery must also be an input to the optimization algorithm. Bernatowicz *et al.*[38] first considered such a routine and they were able to obtain acceptable results assuming that the combination of breathing and delivery matched between planning and treatment. This is an unrealistic assumption as both of these time-dependent processes can vary greatly during treatment and between fractions. Engwall *et al.*[39] took the next step by considering the variation in patient breathing in a robustness framework such that each robust scenario had a different patient-breathing period. They also used Monte Carlo simulation (MC) for their dose calculation, which is important in general for proton therapy[40] and in particular for the thorax geometry.[34,41,42] These were important first steps, but they fall short of a comprehensive solution as this assumes that treatment begins in the same part of the breathing cycle with each treatment, that the extraction of protons from the accelerator has no uncertainty, and that setup and range uncertainties are negligible.

This work presents a proof-of-concept study of a 4D robust optimization methodology for reducing the interplay effect without assuming synchronization, the exact repeatability of treatment, or fraction averaging. This was done by broadening the robust optimization framework to include both the standard physical uncertainties, due to proton range and patient setup, and temporal uncertainties related to the delivery of the treatment, which includes patient breathing and proton extraction. This is a computationally large problem and, as such, the 4D robust optimizer (4DRO) was developed as an extension to an in-house 3D robust optimizer[12,43] (3DRO) which runs on a cluster with many crucial components boosted by graphics processing units (GPUs). The GPU components included the optimization algorithm itself, an in-house MC dose engine,[44] which, as noted above, is particularly important with the density heterogeneities in the thorax anatomy, and an in-house 4D dose calculator (4DDC)[45] which calculates 4D dynamic dose.[31] The 4DRO was evaluated for three test patients with different treatment sites by considering the robustness and homogeneity of plans generated with both the 4DRO and 3DRO.

## 2. METHODS
### 2.A. 4D data sets

The 4DRO obtained information about the patient geometry and motion by using a 4DCT data set phase-sorted[46] into $N_B = 10$ breathing phases labeled $CT_0, CT_{10}, \ldots, CT_{90}$ for the 0%, 10%, and 90% phases, respectively.[b] The institutional standard for motion cases was to generate an averaged CT image set using all phases and then to use the averaged set for contouring and treatment planning; the 3DRO plans were optimized on this data set mimicking what is current done in the clinical setting. The averaged image set was not used in the 4DRO, as it was not a physical phase occurring in the breathing cycle (*i.e.*, not a truly anatomic representation), with the more-stable near-end-of-exhalation phase[47] $CT_{50}$ used instead for optimization evaluation. To conserve computer memory and increase dose-calculation times, CT slices for all series were down-sampled from a $1.27 \times 1.27$ mm$^2$ to a $2.5 \times 2.5$ mm$^2$ pixel size, while maintaining the native slice thicknesses of 1.5 mm.[48]

In order to evaluate on $CT_{50}$, as well as properly place spots and run the MC on each breathing phase, a structure set was needed on each 4DCT phase. The structure set from the averaged CT was transferred to each phase with deformable image registration (DIR) using the MIM software package (version 6.8.3, MIM Software Inc., Cleveland, OH). An exception was the "external" contour, which encompassed the entire patient body, treatment table, and immobilization device plus an additional margin. The external was transferred rigidly to each phase so that the MC would use the same-sized geometry on each phase. MIM was also used for DIR between $CT_{50}$ and the other phases to generate deformation vector fields (DVFs) used in 4D dose accumulation (see Sec. 2.C.3). The DICOM inputs to the 4DRO were then $N_B$ CT series, $N_B$ structure sets, and $(N_B - 1)$ DVF registration files.

## 2.B. Robust scenarios

The robust scenarios for the 3DRO were defined based on eight patient geometries dictated by physical uncertainties. The input CT data were used as the nominal geometry with the $N_P = 8$ robust geometries defined by shifting the nominal by $\pm 3$ mm in the $x$, $y$, and $z$ dimensions, respectively, to simulate set-up error, and by scaling the voxel densities by $\pm 3\%$ to simulate the proton range uncertainty.

The nominal and robust scenarios in the 4DRO were constructed by combining one of the above patient geometries with a manifestation of the treatment delivery as dictated by temporal uncertainties, as described below. The treatment deliveries were simulated using a beam-delivery simulation which was a component of the 4DDC. The result of the beam-delivery simulation was subject to uncertainties in the mean length during the patient's breathing cycle,[c] the breathing phase at the start of treatment, and operation of the synchrotron accelerator (characterized by measured parameters).[45] A nominal delivery was arbitrarily defined as having a starting phase of $CT_0$, a mean breathing period of 5.0 s, and using mean synchrotron-characterization parameters. The nominal delivery was used with the nominal geometry to define the nominal scenario and with each of the above robust geometries to define the physical-uncertainty robust scenarios. Five robust deliveries were defined by changing the nominal delivery by having a starting phase of $CT_{30}$ or $CT_{50}$, $\pm 1$ s change in mean breathing period, and using a variation of the synchrotron characterization parameters (based on measured uncertainties). The $N_T = 5$ temporal-uncertainty robust scenarios each used the nominal patient geometry and one of the above deliveries. The 4DRO thus considered $N_P + N_T + 1 = 14$ total robust scenarios compared to the $N_P + 1 = 9$ total scenarios in the 3DRO.

The beam-delivery simulation determined the spot and breathing-cycle timing for the delivery of a single fraction. Additional fractions were not explicitly considered since fraction averaging was not assumed in this work (see Sec 1) and a uniform final dose was achieved by optimizing such that individual fractions were uniform.

## 2.C. Robust optimization

### *2.C.1. Setup*

In addition to the DICOM files described in Sec. 2.A, the inputs for both 3D and 4D optimizations were: gantry angles, dose constraints for all structures with a priority and robustness directive per structure, physical robust-scenario descriptions with a priority per scenario, and temporal robust-scenario descriptions with a priority per scenario (4DRO only). The optimization algorithm used considered all scenarios (Sec. 2.C.4) and the robustness directive for each constraint indicated whether to consider constraint violations on all scenarios or just the nominal scenario. Much of this flexibility was developed as part of the 3DRO and was not used for the tests presented here where only the target constraints were considered in all scenarios, only the highest priority constraints were used, and all scenario priorities were equal. Initial parsing of the 4DCT, structure sets, and associated DVF DICOM files was performed with a PYTHON script which reformatted necessary information as inputs to the optimization framework (which consisted of a number of C++ and CUDA scripts).

### *2.C.2. Calculation of MC dose influence maps*

A general overview of the 4D optimization algorithm is given by the flowchart in Fig. 1. The starting step was to use the MC to find potential spots[43] whose Bragg peaks were within the clinical target volume (CTV) of each field under any physical uncertainty scenario and, for the 4DRO, on any breathing phase. The scanning target volume (STV) was defined as the union of the above plus a one-spot margin (∼5 mm).

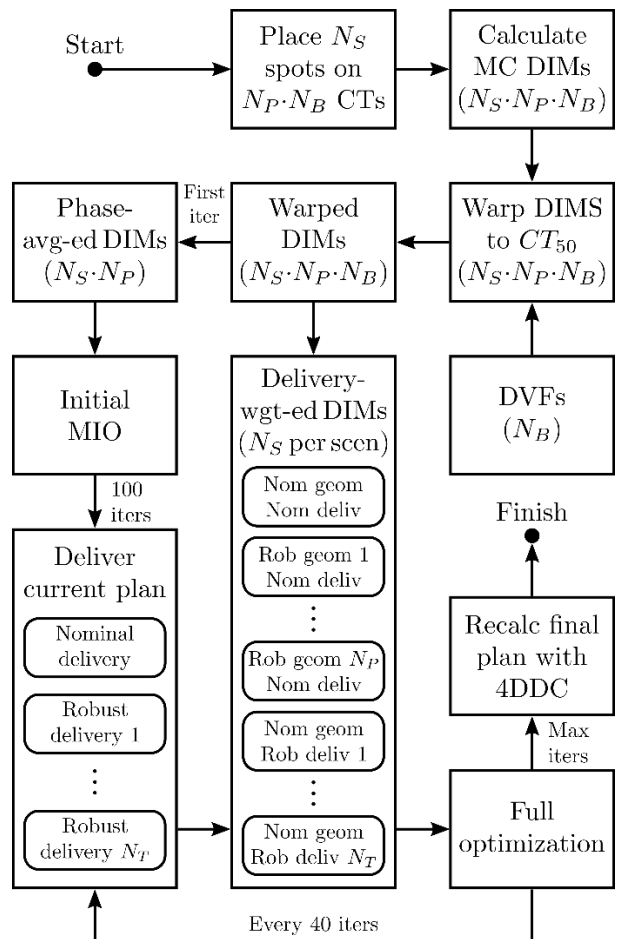

Fig. 1 Flowchart showing the general overview of the 4DRO. Values given in parentheses for a step indicate how many objects were involved in that process, *e.g.*, there were $N_S \cdot N_P \cdot N_B$ DIMs calculated with the MC in the second step. Abbreviations — MC: Monte Carlo, DIM: dose influence map, DVF: deformation vector field, avg-ed; averaged, wgt-ed: weighted, 4DDC: 4D dose calculator, MIO: multi-image optimization, nom: nominal, rob: robust, geom: geometry, deliv: delivery, iter: iteration, $CT_{50}$: 50% breathing phase CT series, $N_S$: number of spots, $N_B$: number of breathing phases, $N_P$: number of physical scenarios, $N_T$: number of temporal scenarios.

Next, 50,000 protons were simulated with the MC for each spot in each physical scenario and breathing phase (4DRO only) to calculate the dose-influence maps (DIMs) for each spot. The MC utilization was slightly different when calculating DIMs compared to placing spots or doing a final recalculation (Sec. 2.C.5). The first difference was compiling the MC as a C++ library such that it could be included directly by the master optimization C++ script. The direct inclusion of the MC avoided writing the DIMs to disk and subsequently increased efficiency.[48] The second difference was that the tails of the DIMs were truncated to conserve memory on the GPU cards[43]; voxels with dose <0.3% of the maximum dose in the target were dropped from each DIM.

### *2.C.3. Dose calculation in 4D*

In the optimization algorithm (see below), each spot $i$ was assigned a fluence weight $\omega_i^k$ which was updated with each iteration $k$. Since the fluence weights were not breathing-phase dependent, the DIMs from the different breathing phases needed to be combined before optimization; this also had the computational advantage that the DIM footprint on the GPUs during optimization was the same as for the 3DRO and thus the optimization logic did not need to be further altered. Combining breathing-phase doses for each spot was performed in two steps, the first of which was to warp the DIMs from non-$CT_{50}$ breathing phases to $CT_{50}$ according to the input DVFs. The warping was performed immediately following the MC with the dose-accumulation component of the 4DDC using dose-interpolation mapping.[45] After this warping, all further calculations for optimization were performed on $CT_{50}$ with the warped DIMs.

The second step was performing a beam-delivery simulation per temporal scenario to determine how to combine DIMs from different breathing phases. The delivery parameters for the given scenario $s$ along with the current fluence-weights vector $\boldsymbol{\omega}^k$ were used to determine a matrix of breathing-phase weights $\boldsymbol{\phi}^{sk}$ which gave the percentage of each spot delivered in each breathing phase. For a given scenario and iteration, the delivery-weighted dose $M_{ij}^{sk}$ in voxel $j$ (on $CT_{50}$) from spot $i$ was then computed as

$$M_{ij}^{sk} = \sum_b \phi_{ib}^{sk} m_{ij}^{sb}, \tag{1}$$

where the summation is over breathing phases and $m_{ij}^{sb}$ is the dose from spot $i$ in voxel $j$ from breathing phase $b$ in scenario $s$ (note that, if $b$ does not correspond to $CT_{50}$, this is the dose from phase $b$ which warped to voxel $j$ on $CT_{50}$). The collection of $M_{ij}^{sk}$'s for all voxels constituted the delivery-weighted DIM for that spot and scenario.

The weights for a given iteration may not have corresponded to a deliverable plan as there were minimum and maximum numbers of monitor units[d] (MUs), $MU_{min} = 0.002$ MU and $MU_{max} = 0.1$ MU, respectively, that the machine would deliver per spot. Before each delivery simulation, the current set of fluence weights were converted to MU and spots reorganized as needed to meet these limits; spots with $MU < MU_{min}$ were dropped (fluence weight set to zero for rest of optimization) and spots with $MU > MU_{max}$ were broken into subspots each with a maximum MU of $MU_{max}$. The spots were then sorted into the proper delivery order for the synchrotron system. The spots with weights above $MU_{max}$ were reconstructed post-beam-delivery simulation to appropriately calculate their breathing-phase weights.

Determining the delivery-weighted dose in Eq. (1) required knowing the fluence weights and, therefore, special consideration was needed for the initial iteration; the initial weights corresponded to an unrealistic plan such that performing a beam-delivery simulation to calculate $\boldsymbol{\phi}^{s0}$ would not be meaningful. Instead, an initial MIO was performed using just the physical scenarios and the phase-averaged dose for each voxel. Formally, this was performed by setting $\phi_{ib}^{s0} = 1/N_B$ for all $i$ and $b$ when $s < N_P$ in Eq. (1). The MIO proceeded with the phase-averaged DIMs for 100 iterations to obtain a more physical fluence-weight distribution before expanding to the full optimization, including the temporal scenarios, dropping min MU spots, and calculating the delivery-weighted DIMs. Re-computing the delivery-weighted DIM after each subsequent iteration would be computationally burdensome so instead $\boldsymbol{\phi}^{sk}$ was used as a constant for 40 iterations between performing beam-delivery simulations. Using the delivery-weighted dose prior to 100 iterations and/or updating $\boldsymbol{\phi}^{sk}$ more frequently led to optimizer convergence issues and suboptimal plans.

The 3DRO also dropped spots below $MU_{min}$ to obtain a deliverable plan. For parity and fair comparison, the spot-dropping logic was the same as in the 4DRO; spots were dropped every 40 iterations following an initial 100 iterations.

### 2.C.4. Robust optimization engine

Multi-field robust optimization was performed using the all-scenario robust optimization method, which has been shown to have superior performance over worst-cast robust optimization strategies.[12] The cost function in the all-scenario algorithm $F(\boldsymbol{\omega}^k)$, for a given iteration $k$, was formulated as

$$F(\boldsymbol{\omega}^k) = \sum_{s} c^s P^{sk} \left\| \boldsymbol{a} \odot \boldsymbol{b} \odot (\boldsymbol{D}^{0k} - \boldsymbol{M}^{sk}\boldsymbol{\omega}^k) \right\|^2, \quad (2)$$

where the summation is over all scenarios, $c^s$ is the input scenario priority, $P^{sk}$ is the DVH-based scenario weight (see below), $\boldsymbol{a}$ is a vector of weights between 0 and 1 indicating structure priorities, $\boldsymbol{b}$ is a vector of structure robustness directives, $\boldsymbol{D}^{0k}$ is a vector of desired doses based on the DVH constraints (see below), $\boldsymbol{M}^{sk}$ is the dose influence matrix for the given scenario, $\boldsymbol{\omega}^k$ is the vector of fluence weights, $\odot$ signifies element-wise multiplication, and $\|\boldsymbol{x}\|$ is the magnitude of vector $\boldsymbol{x}$. The number of rows in $\boldsymbol{M}^{sk}$ was the number of voxels in all structures with constraints and the number of columns was the number of potential spots, *i.e.*, a single column corresponded to a DIM computed in Secs. 2.C.2– 2.C.3 for that spot. In the 3DRO, the influence matrix was iteration independent whereas in the 4DRO it was updated in accordance with Eq. (1). The number of rows in vectors $\boldsymbol{a}$, $\boldsymbol{b}$, and $\boldsymbol{D}^{0k}$ were the number of voxels in the structures with constraints with the values of a given element being those corresponding to the structure to which the voxel belongs (see Sec. 2.C.1 for the values used for $\boldsymbol{a}$, $\boldsymbol{b}$, and c for the present work). For the initial iteration, the fluence weight for each spot was $\omega_i^0 = \text{MU}_{\text{min}}$.

The fluence weights were updated with each iteration, accounting for all scenarios, as[12]

$$\omega_i^{k+1} = \omega_i^k \left\{ \sum_{j,s} \left[ \frac{D_j^{0k}}{d_j^{sk}} \left(M_{ij}^{sk}\right)^2 a_j b_j c^s P^{sk} \right] \Big/ \sum_{j,s} \left[ \left(M_{ij}^{sk}\right)^2 a_j b_j c^s P^{sk} \right] \right\}, \quad (3)$$

where $\boldsymbol{d}^{sk} = \boldsymbol{M}^{sk} \boldsymbol{\omega}^k$ and the scenario-weighting factor was calculated as

$$P^{ks} = \frac{1 + \sum_{m;u} \max[0, \text{DVH}_m^{sk}(d_m) - V_m] + \sum_{m;l} \max[0, V_m - \text{DVH}_m^{sk}(d_m)]}{\sum_{s'} \{1 + \sum_{m;u} \max[0, \text{DVH}_m^{s'k}(d_m) - V_m] + \sum_{m;l} \max[0, V_m - \text{DVH}_m^{s'k}(d_m)]\}}, \quad (4)$$

where $m$ indexes over constraints which are either upper ($u$) or lower ($l$) bounds, respectively, $(d_m, V_m)$ are prescribed dose-volume histogram (DVH) points for constraint $m$, and $\text{DVH}_m^{sk}(d_m)$ is the fractional volume of the structure with constraint $m$ containing a dose greater than $d_m$ in scenario $s$ at iteration $k$. In essence, a scenario's contribution to fluence-weight updating and the cost function was weighted by the constraint violation in that scenario.

While the prescribed DVH points were a global constant, the desired doses $\boldsymbol{D}^{0k}$ used in Eqs. (2) and (3) were periodically updated to avoid unnecessary penalties during the optimization.[12,43] The updating was based on the current doses in the nominal scenario such that if no constraints were violated for a given structure, then $D_j^{0k} = d_j^{0k}$, otherwise

$$D_{j;u}^{0k} = \min_{m}\left(\max\left\{0, d_j^{0k} - [\text{DVH}_m^{0k}(d_m) - V_m]\left[\frac{d_m - d_j^{0k}}{\text{DVH}_m^{0k}(d_j^{0k}) - \text{DVH}_m^{0k}(d_m)}\right]\right\}\right) \quad (5)$$

$$D_{j;l}^{0k} = \max_{m}\left(\min\left\{d_u, d_j^{0k} - [\text{DVH}_m^{0k}(d_m) - V_m]\left[\frac{d_m - d_j^{0k}}{\text{DVH}_m^{0k}(d_j^{0k}) - \text{DVH}_m^{0k}(d_m)}\right]\right\}\right), \quad (6)$$

where the two expressions determined the desired doses depending on whether an upper ($u$) or lower ($l$) constraint $m$ was violated, respectively, and $d_u$ is the lowest violated upper-dose constraint in the target.

### *2.C.5. Final dose calculation*

The optimization terminated when a maximum number of iterations was reached. A final, high-fidelity, dose recalculation was then performed with the optimized plan using the 4DDC for both 3D and 4D optimizations. For both the 3DRO and 4DRO final plans, the recalculation was run on all physical and temporal scenarios to give a collection of final dynamic doses which were converted to DICOM files with another PYTHON script. Using the 4DDC and temporal scenarios with the 3DRO final plans allowed assessment of plan degradation with motion. All final doses were thus dynamic doses evaluated on $CT_{50}$ where DVHs were then computed on the GPU using the same parallel DVH-calculating kernels as were used during the optimization.

### **2.D. 3DRO and 4DRO Comparison Study**

To compare the performance of the 4DRO to that of the 3DRO, plans were retrospectively generated for three test patients with differing disease sites, characterized in Table 1, using the same targets and constraints in both optimizers. The patients were treated clinically with a phase-gated plus repainting delivery, however, the optimizations presented here did not use either delivery modification to assess how well the 4DRO performed as the only motion-mitigation technique. DVH bands for constrained structures as well as DVH metrics for the targets were calculated using the final dynamic doses, from all scenarios, on $CT_{50}$. The DVH metrics assessed target coverage through $D_{98}$ and the interplay effect through the homogeneity index $\text{HI} = D_2/D_{98}$, where $D_x$ is the dose to $x\%$ of the volume.

| Characteristic | Esophagus | Liver | Lung |
|---|---|---|---|
| Total Prescription [Gy] | 50, 45 | 58.05 | 60, 54 |
| Planned Fractions | 25 | 15 | 30 |
| Constraints | 6 | 10 | 5 |
| Fields | 2 | 3 | 4 |
| Target volume [cm$^3$] | 310±7 | 860±16 | 658±4 |
| Tumor motion [mm] | 10.4 | 10.2 | 10.5 |

Table I Characteristics of the three test cases used for 4DRO evaluation against 3DRO. The esophagus and lung cases had two CTVs each at higher and lower prescription levels, respectively. The constraints row indicates the number of DVH points given to the optimizer (the highest-priority constraints) and includes the target constraints. The target volume is given as $\mu \pm \sigma$ calculated from the volumes of the lowest-dose CTV on all ten breathing phases. Tumor motion was estimated as the maximum displacement, between any two phases, of the gross target volume's centroid.

## 3. RESULTS

The dynamic doses from the nominal scenario for both optimization strategies are given in dose wash in Fig. 2 overlaid on the $CT_{50}$ geometry with the highest prescription CTV (see Table I) contoured. For the slices shown (centroid of CTVs), all cases showed improvements with the 4DRO. In the esophagus case, the homogeneity in the target was noticeably improved. For the liver case, the target coverage was improved, particularly in the posterior region of the target. The lung case demonstrated improved coverage (particularly in the inferior region), improved homogeneity, and reduction of overdosed areas.

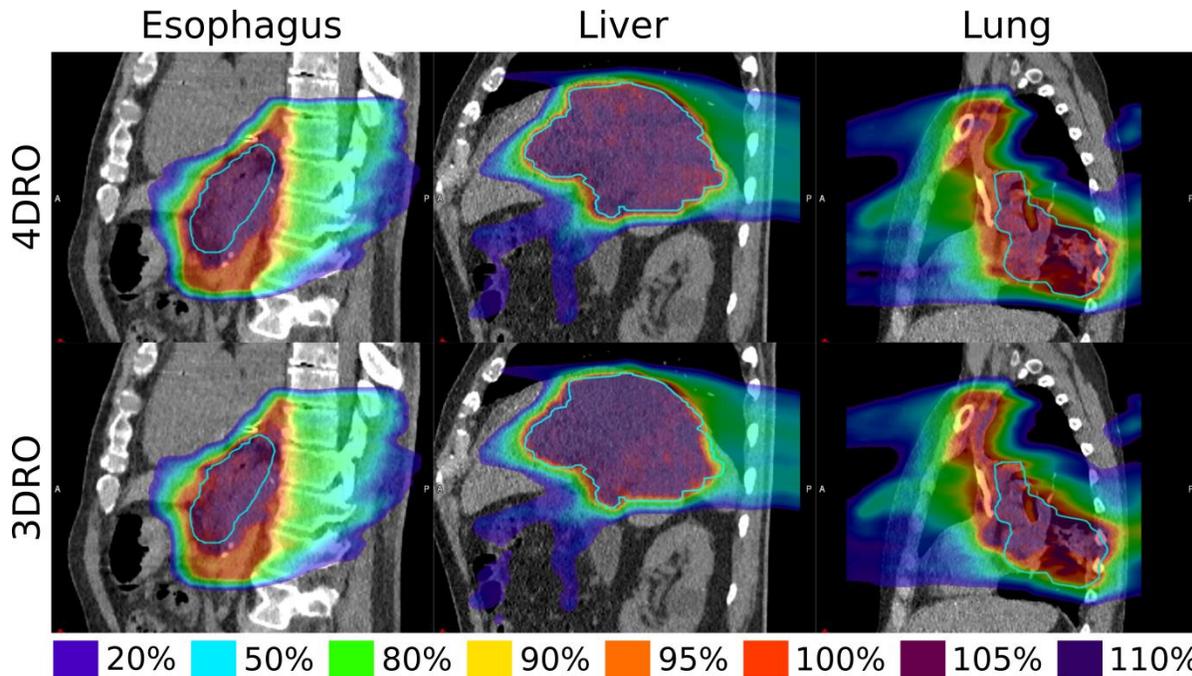

Fig. 2 Dose-wash distributions for the nominal scenario's final doses from the 4DRO (top) and 3DRO (bottom) overlaid on $CT_{50}$ for each of the three cases. Slices shown are from the centroid location of the highest prescription CTV (cyan contour) for each case. The percentages of these prescriptions represented by each dose-wash color are given at the bottom. Noticeable differences between the 4DRO and 3DRO are the homogeneity in the target for the esophagus (left) and lung (right) cases and target coverage for the liver (center) and lung cases, particularly in the inferior and posterior CTV regions, respectively.

DVH bands for structures constrained in the optimizations are presented in Fig. 3 for all test cases. The bands were generated from the DVHs from all 14 scenarios and give the minimum/maximum envelope of the scenarios' curves; the DVHs from the nominal scenarios are also indicated. Most notably, the higher prescription CTVs had improved coverage with the 4DRO, though, for the liver and lung cases the 4DRO target coverage still fell short from the clinically desired result. The doses to the OARs were generally similar, showing perhaps a slight trend towards higher doses observed with the 4DRO results.

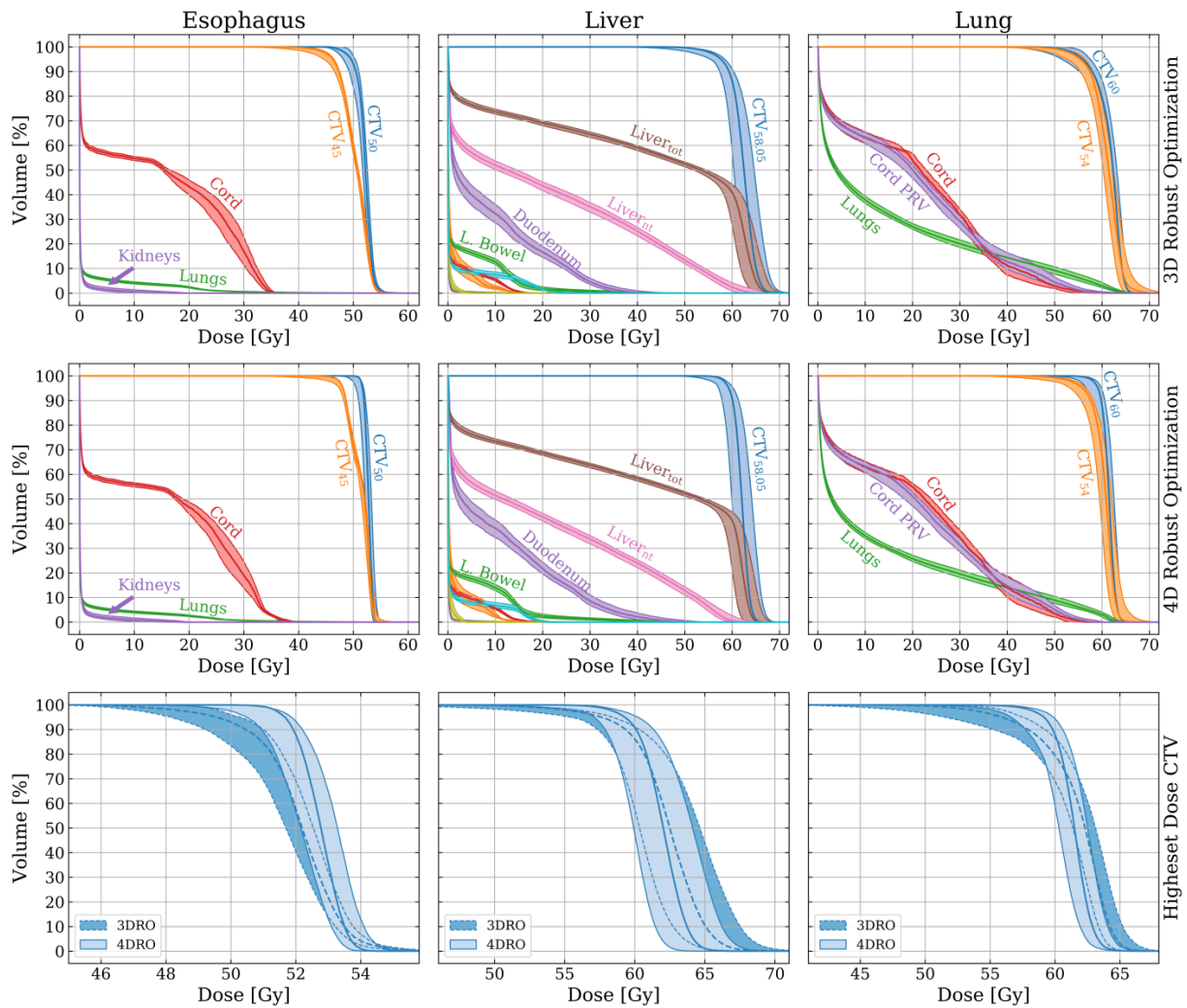

Fig. 3 DVHs for all constrained structures for each of the esophagus (left), liver (center), and lung (right) test cases with optimizations performed with the 3DRO (top) or 4DRO (middle). Regardless of the optimization engine, final doses were calculated using the 4DDC on all 14 scenarios with the 4D dynamic doses evaluated on $CT_{50}$. The bands around each structure give the minimum/maximum envelop of the DVHs for that structure amongst the scenarios; the central curves indicate the DVHs from the nominal scenario. Target coverage and homogeneity are noticeably improved by the 4DRO for the higher-dose CTVs in each case as emphasized by an overlay in the bottom row. The unlabeled bands for the liver case are, given in decreasing volume order at 10 Gy, for the stomach, large bowel, cord PRV, esophagus, and kidneys (the esophagus and kidneys overlap). Abbreviations — $CTV_x$: clinical target volume with prescription of $x$ Gy, $Liver_{tot}$: total liver, $Liver_{nt}$: non-target liver, L. Bowel: large bowel, Cord PRV: spinal cord planning organ at risk volume.

The doses to the targets were more quantitatively explored as given in Fig. 4, which compares DVH metrics between the 3DRO and 4DRO final doses, again using the 4D dynamic doses calculated under all 14 scenarios. The five targets across all three cases (see Table I) were assessed for coverage and homogeneity through $D_{98}$ and HI $= D_2/D_{98}$, respectively. The percent difference between the median levels for both indices, as well as the range amongst the scenarios, is given in Table II. For all targets, the HI distributions moved closer to unity (HI = 1 indicates a homogenous distribution), with an average decrease of 6.0%, and the ranges of the distributions decreased, by an average of 52.5%, with the 4DRO. Similarly, the $D_{98}$ distributions moved towards higher doses, by 4.9% on average, with a decreased average range of 35.0%. Notably, the distributions for the esophagus targets were below the prescription levels with the 3DRO while being above the prescription levels with the 4DRO. The target coverage for the liver and lung targets improved with the 4DRO, as the distributions moved to higher doses, though they were all still below their respective prescription levels. The global decrease in distribution range for both metrics and all targets indicates the 4DRO achieved overall more robust plans.

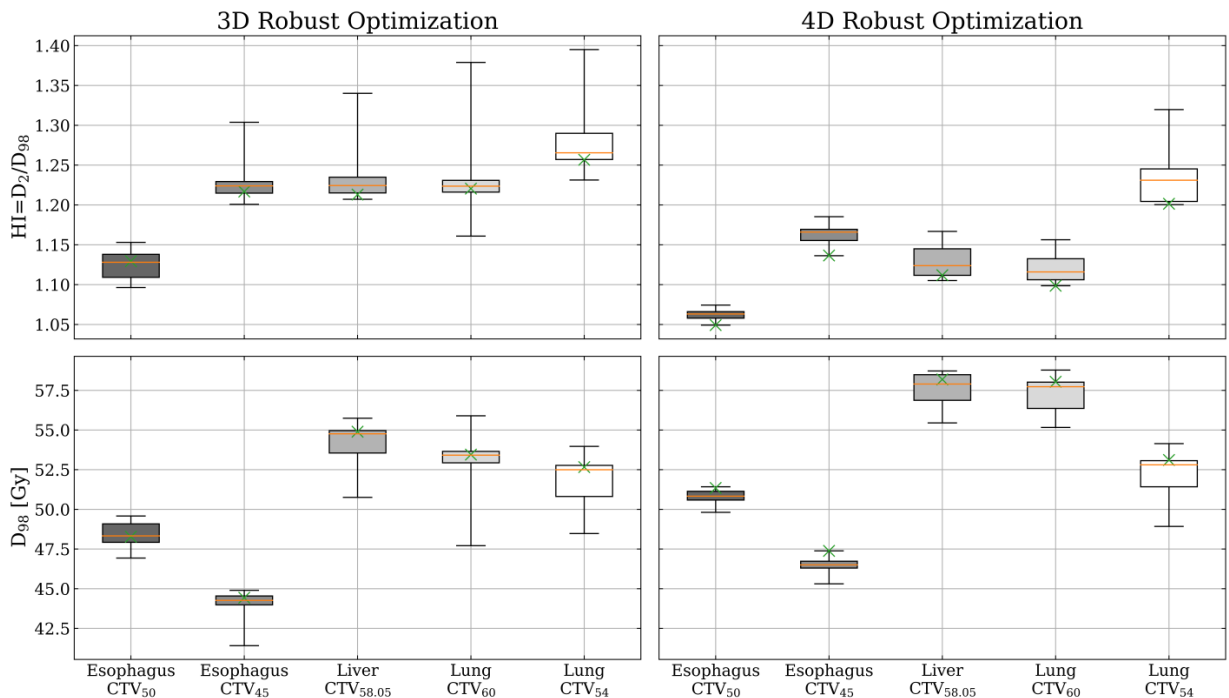

Fig. 4 Target homogeneity (top, assessed with HI = $D_2/D_{98}$) and coverage (bottom, assessed with $D_{98}$) for targets across all test cases for optimizations from the 3DRO (left) and 4DRO (right). Box plots are based on the 4DDC-evaluated doses on $CT_{50}$ for all 14 scenarios; whiskers encompass the entire range of the distribution, the horizontal lines within the boxes give the median positions, and the nominal scenario's values are given by crosses. For all targets, the HI distributions were closer to unity with the 4DRO results which indicates a more uniform (and desirable) distribution that is less affected by the interplay effect. The $D_{98}$ distributions for all targets were shifted to higher dose with the 4DRO for all targets, though they did not reach the prescription levels for the liver and lung targets. The ranges of the distributions were also generally tighter, indicating a more robust plan, with the 4DRO. Abbreviations — $CTV_x$: clinical target volume with prescription of $x$ Gy.

| Case | Target | Median $D_{98}$ | Range $D_{98}$ | Median HI | Range HI |
|---|---|---|---|---|---|
| Esophagus | $CTV_{50}$ | 5.18% | -39.28% | -5.76% | -55.67% |
| | $CTV_{45}$ | 5.05% | -40.59% | -4.74% | -52.48% |
| Liver | $CTV_{58.05}$ | 5.73% | -34.42% | -8.21% | -53.63% |
| Lung | $CTV_{60}$ | 8.08% | -55.84% | -8.80% | -73.51% |
| | $CTV_{54}$ | 0.59% | -4.96% | -2.72% | -27.15% |

Table II Percent difference in the median and range of $D_{98}$ and HI of the 4DRO plans compared to the 3DRO plans for all targets for the three test patients. The range was calculated as the difference between the maximum and minimum values between all 14 scenarios and a decrease in range indicates increased robustness. Median values are the percent difference between the median scenarios in the respective distributions.

## 4. DISCUSSION

The results from Sec. 0 show that the 4DRO achieved more robust plans than the 3DRO as indicated by a general narrowing of the DVH bands and improved target homogeneity and coverage via and HI and $D_{98}$. The liver and lung results indicate that the 4DRO alone would not be sufficient for all cases in mitigating the effects of motion (a similar conclusion was found by Engwall *et al.* [39]) and, *e.g.*, respiratory-phase gating and/or repainting would still be required as with the 3DRO plans. The system described here could handle this scenario as the 4DDC was built to manage both repainting ("maximum-MU"-based repainting, implementation as described by Gelover *et al.*[21]) and phase gating internally and this capability was inherited by the 4DRO. In such cases, the 4DRO could still be of benefit as it may allow less restrictive uses of the alternative delivery methods, *e.g.*, using only phase gating, only repainting, or loosening their individual constraints (increasing the gating duty cycle, higher repainting max MU, etc.). Any of these changes would decrease treatment times, but further study with a larger patient cohort would be required to indicate such changes.

A slight trend towards higher maximum OAR doses with the 4DRO plans was observed in Fig. 1, however, the importance of this must be considered with regard to the constraints given to the optimizers for these OARs. For the cord in the esophagus case, the 4DRO gave a higher maximum dose, with $D_{max}$ = 35.9 Gy for the nominal scenario compared to 34.1 Gy in the 3DRO plan, but the constraint was $D_{max}$ < 45 Gy which was met by both final plans. Another example was the duodenum in the liver case where the 3DRO band terminated at ~40 Gy while the 4DRO band extended higher. The constraint was for <0.5 cm³ of the volume to receive 45 Gy, corresponding to <0.8% of the volume, and this was still met by the 4DRO with the nominal scenario having 0.7% of the volume at ≥45 Gy. Thus, the 4DRO still met the constraints for these OARs, however, it was evidently not able to push the desired dose (see Sec. 2.C.4 and Eqs. (5) and (6)) as low as the 3DRO in some cases (although the target doses for the 4DRO plans were closer to the prescribed doses than those of the 3DRO plans).

There were several general 4D-related uncertainties which would affect the 4DRO. The largest was the reliance on a single 4DCT image set by assuming it represented the patient motion during treatment. Although the 4DRO simulated irregularities in the length of the breathing cycle (by randomly sampling from a distribution per respiratory cycle), it effectively assumed the amplitude of the breathing motion was identical to that of when the 4DCT exam occurred. This could be addressed in the future by modifying the calculated DVFs to change the amplitude or by using a more general motion model fit to multiple 4DCT scans.[49] Additionally, studies have shown that 4D doses can be affected by the type of 4DCT sorting[46] as well as the number of phases used[50]; these studies include 4D optimization approaches which interpolated to >10 phases.[38,51] This study used MIM as the DIR engine and ignored any uncertainties associated with the image registration, however, a 5–10% variation in target $D_{95}$ has been reported between various DIR algorithms[52] and this could impact these results.

The characterization and use of the beam-delivery simulation also included potential uncertainties and limitations. The characterization parameters were measured on a synchrotron of the same design and configuration as our institution's system, however, no two systems are identical and day-to-day variations with the systems have been observed. Using these parameters in both the 4DRO and 4DDC was thus an approximation for our system; including uncertainties on these parameters as a temporal scenario was motivated by this variation, but the exact delivery of a given fraction could not be precisely predicted which led to inherent uncertainty. The nominal delivery was used for all physical scenarios and could have had a greater effect on the optimization than the robust deliveries. While the patient-breathing period and synchrotron characterization could have meaningful nominal values, the initial breathing phase would not and thus the arbitrary selection of $CT_0$ could have affected the final result. Additionally, the choice of performing MIO for 100 iterations and then updating the delivery every 40 iterations thereafter could have affected the final results. These values were chosen after a rough scan of potential values and likely could be optimized further.

Future work could focus on two areas, validation and characterization of the current implementation and pushing towards more advanced avenues of 4D optimization. Studies to validate the 4D components of the 4DRO against measurements are ongoing, including comparing against accelerator log files and motion-phantom interplay measurements. The performance of the 4DRO on a larger patient cohort would better characterize the potential benefits of the 4DRO and inform its future use. The 4D optimization engine itself could be improved by giving the algorithm more direct control over the potential spots. This could include preferring spots whose DIMs are less affected by motion, the pattern of spots considered, and the order in which spots would be delivered. Manipulating spot delivery has been shown to improve interplay reduction,[53] and thus, allowing the optimizer to adjust these spot-specific quantities has the potential to improve the 4DRO's ability to do the same.

## 5. CONCLUSIONS

This article presented a proof-of-concept study for a Monte-Carlo based 4D robust optimization methodology which had the goal of generating treatment plans that increased target coverage, reduced interplay effects, and improved plan robustness for IMPT treatments affected by patient respiration. Most notably, the 4DRO algorithm did not depend on any synchronization of breathing and delivery or on their repeatability with each fraction through the direct inclusion of the interplay effect, and the temporal uncertainties that affect it, into the optimization process. As interplay effects were directly considered in the optimization, the 4DRO also did not need to assume fraction averaging. The 4DRO was compared to 3D-robustly-optimized plans evaluated under motion for test cases and the 4DRO was found to generally improve target coverage and homogeneity. Future studies include the validation and characterization of the current system along with improving the algorithm to further allow explicit spot-specific manipulations.

## ACKNOWLEDGMENTS

This work was supported in part by a grant from Varian Medical Systems.

## CONFLICTS OF INTEREST

The authors have no relevant conflicts of interest to disclose.

Proton Pencil Beam Scanning by Spot-Adapted Layered Repainting Evenly Spread out Over the Full Breathing Cycle. *Int J Radiat Oncol*. 2018;100(1):226-234. doi:10.1016/j.ijrobp.2017.09.043

---

[a] Author to whom correspondence should be addressed. Electronic mail: pepin.mark@mayo.edu.

[b] 4DCT imaging and phase-sorting performed with commercial equipment and software from Siemens Healthineers (Erlangen, Germany).

[c] During the simulation, the length of a breathing phrase was taken as $1/N_B$ times a value randomly sampled from a normal distribution with an input mean and a standard deviation of 0.5 s. A new random value was sampled with each new breathing phase.

[d] 1 MU corresponds to approximately $4 \times 10^8$ to $9 \times 10^8$ protons and is dependent on beam energy.